\documentstyle[aps,prc,preprint,epsf,twoside]{revtex}
\newcommand{\braket}[2]{\left\langle{#1}\right.\left|{#2}\right\rangle}

\newcommand{\bramket}[3]{\left\langle\,{#1}\,\left|\,{#2}\,
            \right|\,{#3}\,\right\rangle}

\newcommand{\intp}[1]{\int\frac{d^{3}{#1}}{(2\pi)^{3}}}
\newcommand{\intpp}[1]{\int\frac{d^{4}{#1}}{(2\pi)^{4}}}
\newcommand{\bqn}{\begin{eqnarray}}
\newcommand{\eqn}{\end{eqnarray}}
\newcommand{\nn}{\nonumber\\}
\newcommand{\Bqn}{\begin{eqnarray}}
\newcommand{\Eqn}{\end{eqnarray}}
\newcommand{\ov}{\overline}
\newcommand{\eps}{\epsilon}

\newcommand{\zweivek}[2]{ \left( \begin{array}{c} #1 \\ #2
                               \end{array} \right) }

\newcommand{\half}{\frac{1}{2}}

\newcommand{\deriv}[1]{\frac{\partial}{\partial #1}}

\begin{document}

\title{Electromagnetic form factors of the nucleon in a covariant
  diquark model}
\author{V. Keiner}
\address{Institut f\"ur Theoretische Kernphysik,\\
         Universit\"at Bonn, Nussallee 14-16, 53115 Bonn, FRG}
\date{September 13, 1995}
\preprint{\vbox{Bonn TK-95-23}}
\maketitle

\begin{abstract}
We present a simple covariant constituent diquark-quark model for the
nucleon. The nucleon is assumed to be composed of a
scalar diquark and
a quark which interact via a quark exchange. Starting from the
Bethe-Salpeter equation, the instantaneous approximation leads
to a diquark-quark Salpeter equation. In the Mandelstam
formalism, the electromagnetic form factors of the nucleon
are calculated for momentum transfers up to $q^2 = - 3 \; (\mbox{GeV/c})^2$.
A remarkable description of the experimental data is obtained.
Especially, the model gives nearly the right values for the proton and
(negative) neutron charge radii, and a qualitative description of the
magnetic form factors.
\end{abstract}

\vspace{1cm}

\vspace{5cm}
e-mail: {\em keiner@pythia.itkp.uni-bonn.de}

\newpage

\pagenumbering{arabic}

\section{Introduction}
The possible existence of diquarks within the nucleon has been
discussed for a long time. Undoubtly, there is much experimental evidence
for diquarks or diquark correlations/clustering.
For a review on that subject
see e.g. \cite{anselmino,dziembowski,fred}. One main argument for
diquarks is the possible explanation of the negative neutron mean
square charge radius \cite{dziembowski}, which cannot be explained
in a pure SU(6) quark model. \\
On the other hand, of course, the analytical and numerical
treatment of a two-body system
is much less complicated than a three-body calculation.
This is especially the case if one treats the problem
relativistically using a Faddeev-type equation. In the last years,
using the notion of diquarks, much progress has been done in this
field.
Especially, the description of
baryons in the framework of the
Nambu--Jona-Lasinio model \cite{ishii,huang,reinhardt,hanhart} has
been successful.
There, two quarks within the baryon are treated as diquark states
analogous to mesons (both are $\ov 3$ in colour space). The interaction
with the third quark then is modelled by a quark exchange. Lichtenberg
\cite{lichtenberg} proposed this short-range interaction long
ago to split the degeneracy of the (70,0$^+$) and (56,0$^+$) ground state in a
quark model of a SU(6) 21-plet diquark and a 6-plet quark.
If one assumes a diquark clustering in the nucleon,
it is very natural to assume such an interaction. In that case, none
of the identical quarks is distinguished although we have diquark
correlation. In \cite{vogl,weiss,hellstern} masses
and diquark form factors are calculated within the NJL model. Cahill et al.
\cite{cahill} have calculated the masses of scalar and axial-vector
diquarks by using an approximate form of the Bethe-Salpeter equation
for qq bound states. Most of the above references conclude a probable
dominant role of the scalar diquark compared to the axial-vector one. The
parameters used in the above literature will serve us as a guideline.

Any relativistic model of the nucleon should of course reproduce
correctly the electromagnetic form factors of the nucleon for non-zero
momentum transfers.
Meyer \cite{meyer} solves numerically the BS equation for quark-diquark
bound states with a quark exchange potential without, however,
calculating any dynamical observables.
Kroll et al. \cite{kroll,anselminokroll} successfully calculate form
factors of the
nucleon in a diquark model using light-cone distribution
amplitudes. Of course, their model is applicable only at higher momentum
transfers.

In our framework, form factors can be calculated in a
straightforward way once the (covariant) amplitudes are known. We
present a relativistic constituent quark model
of the nucleon where the nucleon is assumed to be a bound state of a
scalar diquark and a quark. Following the motivation given above, the
interaction between the two particles is modelled by the exchange of a
quark. Since we are only interested in the (ground state) nucleons, we
do not incorporate a phenomenological confinement potential. We just
want to show, that, by using only a (short-range) quark exchange
potential, it is
possible to describe the form factors with physically acceptable
parameters. The fundamental equation is the Bethe-Salpeter equation for a
bound state of a scalar particle and a fermion. Assuming an
instantaneous interaction kernel we derive a
Salpeter-type equation. This can be cast in a Schr\"odinger-type
equation. This eigenvalue equation is solved numerically by standard
techniques. The formal covariance of the Salpeter
equation and the Salpeter amplitudes is discussed extensively in
\cite{muenz}. We then use a straightforward method to calculate
the electromagnetic currents in the Mandelstam formalism, analogous to
\cite{muenz} where meson form factors are computed.
Since form factors are directly related to wave functions,
the calculation of them is a crucial test for any model.

This paper is structured as follows. In Sect.\ref{model} we derive
the Salpeter equation for a bound state of a scalar diquark and a quark,
including the normalization of the amplitudes. The assumed
interaction between diquark and quark is outlined in
Sect.\ref{interaction}.
In Sect.\ref{solution} the procedure solving the Salpeter equation is
explained. The computation of the electromagnetic current and of the
resulting form factors
of the nucleon is discussed in Sect.\ref{formfactors}. The results
for the form factors are presented in Sect.\ref{results} and
compared to experimental data. Finally, in Sect.\ref{summary},
a summary is given, and an outlook to an extension of the
model and to possible applications.

\section{The Model}
\label{model}

\subsection{The Bethe-Salpeter equation}

A relativistic bound state of a scalar particle and a fermion with
four momentum $P$ is
described by a Bethe-Salpeter amplitude
\bqn
\chi_P(x_1,x_2) = \bramket{0}{T \phi(x_1) \psi(x_2)}{P} \quad.
\eqn
In momentum space, $\chi_P$ fulfills the following Bethe-Salpeter
equation \cite{BS}
\bqn \label{BSequation}
\chi_P(p) = \Delta_1^F(p_1) S_2^F(p_2)
            \intpp{p'} (-i K(P,p,p')) \chi_P(p')
\eqn
with $\Delta_1^F(p_1) = i/(p_1^2-m_1+i\eps)$ and $S_2^F(p_2) =
i/(\gamma p_2-m_2+i\eps)$ the usual Feynman propagators for
a scalar particle and a quark, respectively, and $-i K$ the
irreducible interaction kernel.
The masses in the above propagators are constituent masses of
the order of a few hundred MeV. For the momenta we set
\bqn
& & p_1 = \eta_1 P+p \;,\; p_2 = \eta_2 P-p \quad;\quad
P = p_1+p_2 \;,\; p = \eta_2 p_1 - \eta_1 p_2 \nn
& &\mbox{with} \quad \eta_1+\eta_2 = 1 \quad.
\eqn
The normalization condition is obtained as usual by
considering the pole contribution of the two-particle (4-point-)
Green's function (see e.g. \cite{nakanishi})
\bqn
G(x_1,x_2,x_1',x_2') = \bramket{0}
{T \phi(x_1) \psi(x_2) \phi^\dagger(x_1') \ov\psi(x_2')}{0}
\eqn
to the bound state:
\bqn
G(P,p,p') \sim \frac{i}{(2 \pi)^4}\frac{\chi_P(p) \cdot \ov\chi_P(p')}
          {2 \omega_P (P^0-\omega_P+i\epsilon)} \quad.
\eqn
$\omega_P = \sqrt{\vec P^2+M^2}$ is the on-shell energy of the bound
state.
This yields
\bqn \label{normal1}
\frac{i}{(2 \pi)^4} \int d^4p \intpp{p'} \ov\chi_P(p) \deriv{P^0}
  [H(P,p,p')]_{P^0=\omega_P} \chi_P(p') = 2 \omega_P
\eqn
with
\bqn
H(P,p,p') = \Delta_1^F(p_1)^{-1} S_2^F(p_2)^{-1} (2 \pi)^4 \delta^4(p-p')
    +i K(P,p,p') \quad.
\eqn
In covariant form, Eq.\ref{normal1} reads:
\bqn \label{normal2}
\frac{i}{(2 \pi)^4} \int d^4p \intpp{p'} \ov\chi_P(p) P^\mu \deriv{P^\mu}
  [H(P,p,p')]_{P^0=\omega_P} \chi_P(p') = 2 M^2
\eqn
Following Salpeter \cite{salpeter} we neglect the time (i.e. energy)
dependence of the interaction kernel by
assuming $K(P,p,p') = V(\vec p, \vec p\,')$
({\em instantaneous interaction}) in the rest frame of the bound
state.
Then, one can easily perform the $p^0$ integration: Define in the c.m. frame
of the bound state
\bqn
\Phi(\vec p) = \left( \int \frac{d p^0}{2\pi} \chi_P(p^0,\vec p) \right)
                      _{P=(M,\vec 0)}  \quad.
\eqn
Then, from Eq.\ref{BSequation} one gets with standard techniques
\bqn \label{salpeter1}
\Phi(\vec p) = \frac{1}{2 \omega_1} \left(
 \frac{\Lambda_2^+(-\vec p) \gamma^0}{M-\omega_1-\omega_2}
+\frac{\Lambda_2^-(-\vec p) \gamma^0}{M+\omega_1+\omega_2}
\right)
\intp{p'} V(\vec p,\vec p\,') \Phi(\vec p\,')  \quad,
\eqn
with $\Lambda^\pm(\vec p) = \frac{\omega \pm H(\vec p)}{2 \omega}$
the usual Dirac projectors.
$H(\vec p) = \gamma^0 (\vec \gamma \vec p + m)$ is the Dirac
Hamiltonian.
%recall the relation
%\bqn
%S^F(p) = i \; \left( \frac{\Lambda^+(\vec p)}{p^0-\omega+i\epsilon}
% +            \frac{\Lambda^-(\vec p)}{p^0+\omega-i\epsilon} \right)
%\gamma^0  \quad.
%\eqn
Define now
\bqn
\Psi(\vec p) & = & \gamma^0 \Phi(\vec p) \nn
W(\vec p, \vec p\,') & = & V(\vec p, \vec p\,') \gamma^0 \quad.
\eqn
Then, we can rewrite Eq.\ref{salpeter1} in a Schr\"odinger-type
equation:
\bqn \label{salpeter2}
({\cal H} \Psi)(\vec p) & = & M \Psi(\vec p) \nn
& = & \frac{\omega_1+\omega_2}{\omega_2} H_2(\vec p) \Psi(\vec p)
      +\frac{1}{2 \omega_1} \intp{p'}
       W(\vec p, \vec p\,') \Psi(\vec p\,') \nn
& = & ({\cal T} + {\cal V}) \Psi (\vec p) \quad.
\eqn
This is the fundamental equation of our model.
$H_2(\vec p)$ is again the
Dirac-Hamiltonian of the quark. If $\Psi$ were a free Dirac spinor,
$\Psi(\vec p) = u(\vec p)$, the kinetic energy term in
Eq.\ref{salpeter2} is just the sum of the kinetic energies of the two
particles: ${\cal T} = \omega_1+\omega_2$.
In a similar way, one obtains the normalization condition for the
instantaneous approximation from Eq.\ref{normal2}:
\bqn
\frac{i}{(2 \pi)^4} \int d^4p \; \ov\chi_P(p) P^\mu \deriv{P^\mu}
  ( \Delta_1^F(p_1)^{-1} S_2^F(p_2)^{-1})
  \chi_P(p') = 2 M^2   \quad.
\eqn
Performing the $p^0$ integration, one ends up with a familiar
normalization of the functions $\Phi(\vec p)$:
\bqn \label{normal3}
\intp{p} (2 \omega_1) \ov \Phi(\vec p) \gamma^0 \Phi(\vec p) = 2 M
\quad.
\eqn
It is easy to see that $\ov \Phi(\vec p) = \Phi^\dagger(\vec p)
\gamma^0$. So apart from the factor $2 \omega_1$, we obtain the usual
normalization of Dirac spinors. In particular, the functions $\Psi$
solving Eq.\ref{salpeter2} have non-negative norm for $M > 0$.
We thus define a {\em scalar product} as follows (compare
\cite{resag} in the case of two fermions):
\bqn
\braket{\phi}{\psi} = \intp{p} (2 \omega_1) \ov\phi(\vec p) \gamma^0
\psi(\vec p) \quad.
\eqn
In order to obtain a pseudo-Hamiltonian ${\cal H}$ which is hermitian
with respect to the above scalar product, the kernel
$W(\vec p, \vec p\,')$ has to fulfill the following condition:
\bqn  \label{hermitv}
W(\vec p,\vec p\,')^\dagger & = & W(\vec p\,', \vec p) \nn
\mbox{or} \quad V(\vec p, \vec p\,')^\dagger & = &
\gamma^0 V(\vec p\,', \vec p) \gamma^0 \quad.
\eqn
In general, taking into account also anti-particle states, the
solutions of Eq.\ref{salpeter2} come in pairs. This can be seen as
follows. Under charge conjugation, the (Bethe-)Salpeter  amplitudes
transform like (see Eq.\ref{charge})
\bqn
\chi_P^{b_1 f_2}(p) & = & S_C \chi_P^{\ov b_1 \ov f_2}(-p)^* \nn
\rightarrow \Psi_{b_1 f_2}(\vec p) & = &
- S_C  \Psi_{\ov b_1 \ov f_2}^*(-\vec p) \quad
\eqn
with $S_C = i \gamma^2$. With $b_1, f_2$ and $\ov b_1, \ov f_2$ we
denote the quantum numbers of the diquark and quark and their
antiparticles, respectively.
With $S_C H_2(\vec p) S_C = - {}^tH_2(-\vec p)$ and the
imposed condition
\bqn \label{chargec}
W(\vec p, \vec p\,') = - S_C
   {}^tW(-\vec p, -\vec p\,') S_C
\eqn
Eq.\ref{salpeter2} results in
\bqn
({\cal H} \Psi_{\ov b_1 \ov f_2})(\vec p) & = &
- M^* \Psi_{\ov b_1 \ov f_2}(\vec p) \nn
& = & \frac{\omega_1+\omega_2}{\omega_2} H_2(\vec p)
      \Psi_{\ov b_1 \ov f_2}(\vec p)
      +\frac{1}{2 \omega_1} \intp{p'}
       W(\vec p, \vec p\,') \Psi_{\ov b_1 \ov f_2}(\vec p\,') \quad.
\eqn

\subsection{The interaction kernel}
\label{interaction}

\begin{figure}
\begin{center}
\input{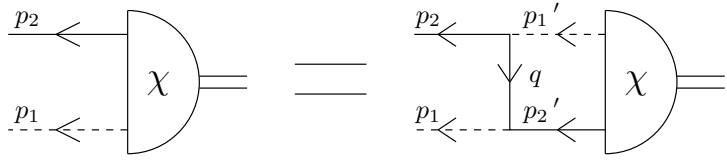}
\end{center}
\caption{The Bethe-Salpeter equation for a bound state of a quark and
  a scalar diquark with a quark exchange interaction}
\label{BSfig}
\end{figure}

The interaction between the scalar diquark and the quark is assumed to
be a quark exchange, see Fig. \ref{BSfig}. The diquark couples
pointlike to the two quarks according to the Lagrangian \cite{meyer}
\bqn
{\cal L}_{int} = -i \; g_s \; \ov\psi_c \; \gamma^5 \;
\tau_2 \; \psi \;\phi^*  \quad.
\eqn
The kernel $K(P,p,p')$ then is
\bqn
K(P,p,p') \chi_P(p') & = & -i \; g_{dqq}^2 \; S_q^F(p+p') \; \chi_P(p') \nn
\rightarrow  V(\vec p, \vec p\,') & = &
-g_{dqq}^2 \frac{1}{\omega_q^2} (-\vec\gamma(\vec p+\vec p\,') + m_q) \quad,
\eqn
with $\omega_q = \sqrt{(\vec p + \vec p\,')^2+m_q^2}$ the energy of
the exchanged quark; set $m_q = m_2$ (mass of the quark).
In the instantaneous approximation we neglect the $p^0$ dependence of
the quark propagator. $g_{dqq}$ is a dimensionless diquark-quark coupling
constant. Obviously, $V(\vec p, \vec p\,')$ fulfills the
relation \ref{hermitv}. In addition, it does not mix states with
different parities since
\bqn
\quad V(\vec p, \vec p\,') & = & \gamma^0 V(-\vec p, -\vec p\,')
\gamma^0
\eqn
and Eq.\ref{chargec} holds for the above
$W(\vec p, \vec p\,') = V(\vec p, \vec p\,') \gamma^0$.

\subsection{Solving the instantaneous diquark-quark equation}
\label{solution}

\begin{figure}
\begin{center}
\leavevmode
\epsfxsize=0.65\textwidth
\epsffile{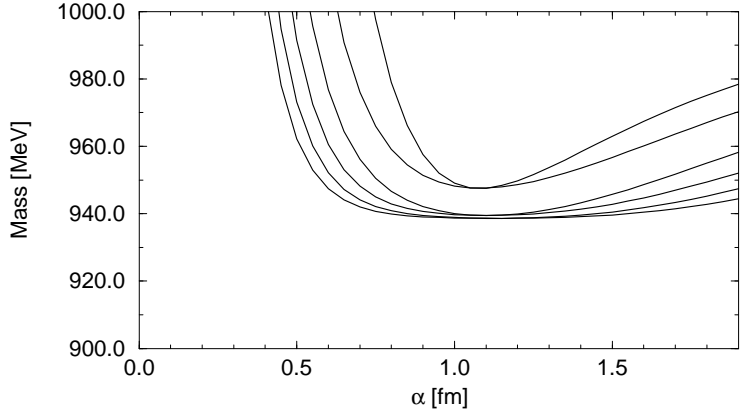}
\end{center}
\caption{The lowest positive eigenvalue of ${\cal H}$
  (Eq.\protect\ref{salpeter2}) as a function of the oscillator parameter
  $\alpha$ for different numbers of basis functions (1 to 6 from top
  to bottom)}
\label{alphaplot}
\end{figure}

Since nucleons have positive parity and spin $\half$ we choose as
basis in Dirac space the following states ($s = \pm \half$):
\bqn \label{basis}
e_s^1(\vec p) & = & \sqrt{W+M_{1 2}} \zweivek{\chi_s}{0} \nn
e_s^2(\vec p) & = & \sqrt{W+M_{1 2}}
\zweivek{0}{\frac{\vec \sigma \vec p}{W+M_{1 2}} \chi_s} \quad,
\eqn
with $M_{1 2} = \frac{m_1 m_2}{m_1+m_2}$ and
$W = \sqrt{\vec p \, ^2+{M_{12}}^2}$.
This ansatz neglects a relative angular momentum between the quark and the
diquark. Both states are multiplied by independent radial functions
which are chosen to be eigenstates of the harmonic oscillator:
${\cal R}^i(p) = (-1)^i \tilde N_{i \; 0}
\sum_{\mu = 0}^i c_\mu^{i \; 0} \; p^{2 \mu}
\mbox{exp}(-\frac{1}{2} \alpha^2 p^2) $.
$\alpha$ is the oscillator parameter.
Eq.\ref{salpeter2} is then diagonalized.
The lowest positive
eigenvalue is interpreted as the nucleon mass. It is obtained
according to the Ritz variational principle by choosing the oscillator
parameter $\alpha$ in the minimum of the curve $M(\alpha)$, see
Fig.\ref{alphaplot}. The coupling $g_{dqq}$ is determined by the condition
$M = 939\;\mbox{MeV/${c^2}$}$. We observe that for a pointlike
coupling of the two quarks to the diquark, there is no stable
solution. Only if one introduces a diquark form factor the minimum of
the $M(\alpha)$ curve converges with increasing basis states, see
Fig.\ref{alphaplot}.  We choose a form factor of the following form:
\bqn \label{diquarkff}
F(\vec q) = \mbox{exp}(-\lambda^2 \vec q \, ^2)
\eqn
and put $\lambda = 0.18 \;\mbox{fm}$, see Sect.\ref{results},
where we study the dependence of our results on $\lambda$.

\section{Electromagnetic form factors of the nucleon}
\label{formfactors}

\begin{figure}
\begin{center}
\input{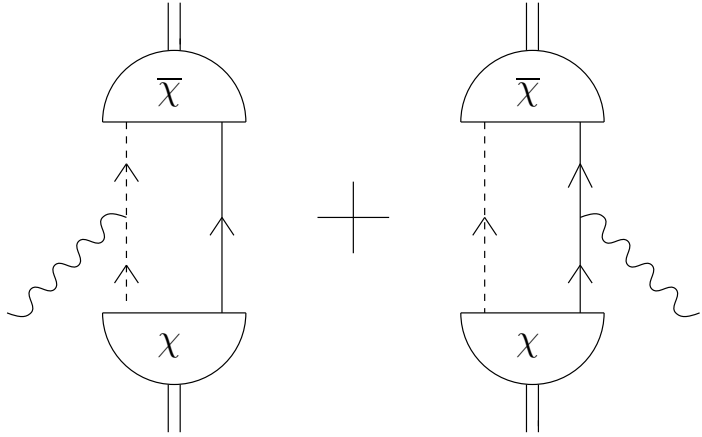}
\end{center}
\caption{The electromagnetic current is the sum of the diquark current
  and the quark current
  (Eqs. \protect\ref{qcurrent1},\protect\ref{dcurrent})}
\label{currentfig}
\end{figure}

In this section we calculate the electromagnetic form factors
of the nucleon in the Mandelstam formalism
\cite{mandelstam,lurie}. In this framework, any current matrix
elements between bound states can be calculated in a covariant way
\cite{muenz}. In our model, the
electromagnetic current coupling to the nucleon is simply the sum of
the diquark- and quark current, see Fig.\ref{currentfig}.
For the coupling to the quark one
obtains (put $\eta_1=\eta_2=\half$)
\bqn \label{qcurrent1}
\bramket{P's'}{j_\mu^{quark}(x=0)}{P s} = e_q \intpp{p}
\ov\chi_{P'}(p+\frac{q}{2}) \gamma_\mu \Delta_1^F(\frac{P}{2}+p)^{-1}
\chi_P(p) \quad.
\eqn
Now introduce the vertex functions as amputated Salpeter amplitudes:
\bqn
\Gamma_P(p) & = & \Delta_1^F(p_1)^{-1} S_2^F(p_2)^{-1} \chi_P(p) \nn
\ov\Gamma_P(p) & = & \ov\chi_P(p) \Delta_1^F(p_1)^{-1} S_2^F(p_2)^{-1}
\quad.
\eqn
{}From Eq.\ref{BSequation} follows in the instantaneous approximation
\bqn \label{vertex}
\Gamma(\vec p) := \Gamma_P(p)|_{P=(M,\vec 0)}
 & = & -i \intp{p'} V(\vec p, \vec p\,') \Phi(\vec p\,') \\
\mbox{and } \ov\Gamma(\vec p) & = & - \Gamma^\dagger(\vec p) \gamma^0 \quad.
\eqn
Then, the rhs. of Eq.\ref{qcurrent1} becomes
\bqn \label{qcurrent2}
& = & e_q \intpp{p} \ov\Gamma_{P'}(p+\frac{q}{2})
S_2^F(\frac{P}{2}-p-q) \gamma_\mu S_2^F(\frac{P}{2}-p)
\Gamma_P(p) \Delta_1^F(\frac{P}{2}+p) \quad.
\eqn
Analogously, for the current coupling to the scalar diquark,
one obtains
\bqn \label{dcurrent}
\bramket{P's'}{j_\mu^{diquark}(0)}{P s}
 = e_d \intpp{p}
\ov\chi_{P'}(p-\frac{q}{2}) S_2^F(\frac{P}{2}-p)^{-1}
(p_1+p_1')_\mu \chi_P(p) \nn
= e_d \intpp{p} \ov\Gamma_{P'}(p-\frac{q}{2})
S_2^F(\frac{P}{2}-p) \Gamma_P(p)
\Delta_1^F(\frac{P}{2}+p) (P+2p-q)_\mu
\Delta_1^F(\frac{P}{2}+p-q) \quad,
\eqn
where we assumed a coupling of the photon to a pointlike scalar
particle. Since the vertex functions $\Gamma(\vec p)$ are only known
in the nucleon rest frame (Eq.\ref{vertex}),
the vertex function of the outgoing nucleon has to be boosted
according to the value of $q^2$, see Eq.\ref{lorentz}.
This covariant treatment of the vertex
functions is very important for high $q^2$. The $p^0$ integration in
Eqs.\ref{qcurrent2},\ref{dcurrent} has the generic form
\bqn
\int_{-\infty}^{+\infty} dp^0 \frac{f(p^0)}
{\prod_{i=1,6}(p^0-p_i^0 \pm i \eps)}
\eqn
with $f(p^0)$ a regular function. The six poles arise from the
denominators of the three propagators. The integral can be split in a
residue part and a principal value integral. To obtain a hermitian
current the latter should vanish.
The $\phi_p$ integration is trivial, since $p$ and
$p'=p \pm \frac{q}{2}$ have the same $\phi_p$. The numerical
treatment is further simplified by expanding the vertex functions in
the basis of  Eq.\ref{basis}:
\bqn
\Gamma_s(\vec p) = \sum_i \;
a_i^1(p) e_s^1 +a_i^2(p) e_s^2(\vec p) \quad ,
\eqn
where $\sum_i$ sums over the radial basis states. \\
We recall that the normalization condition (Eq.\ref{normal3}) equals
the zero-component of the currents at $q^2 = 0$:
\bqn \label{normal4}
\frac{1}{e_q} \bramket{P \; s}{j_0^{quark}}{P \; s}
& = & \frac{1}{e_d} \bramket{P \; s}{j_0^{diquark}}{P \; s}
= 2 M \\
& = & - \intp{p} \frac{1}{2 \omega_1}
\left( \frac{\ov \Gamma(\vec p) \Lambda_2^+(-\vec p)
             \gamma^0 \Gamma(\vec p)}
            {(M-\omega_1-\omega_2)^2}
      +\frac{\ov \Gamma(\vec p) \Lambda_2^-(-\vec p)
             \gamma^0 \Gamma(\vec p)}
            {(M+\omega_1+\omega_2)^2} \right) \nonumber.
\eqn

The nucleon electromagnetic current can be
written in the usual form \cite{bjorken}
\bqn \label{current}
\bramket{P' s'}{j_\mu(0)}{P s} = e \;
\ov u_{s'}(P') \left[\gamma_\mu \; F_1^N(q^2)
+\frac{i \sigma_{\mu \nu} q^\nu}{2 M} \kappa_N \; F_2^N(q^2)\right]
u_s(P) \quad,
\eqn
with $F_1^N$ and $F_2^N$ being independent form factors.
At $q^2 = 0$ we have
\bqn
F_1^p(0) = \frac{1}{e}
           \frac{ \bramket{P \; s}{j_0^{quark}}{P \; s}
                + \bramket{P \; s}{j_0^{diquark}}{P \; s}
                }
                {2 \; M} = 1 \quad,
\eqn
which thus serves as a check of the numerical evaluation of the
currents at zero momentum transfer, i.e. of the normalization.
Usually,
one defines the following form factors:
\bqn
G_E^N(q^2) & = & F_1^N(q^2)+\frac{\kappa_N q^2}{4 M^2} F_2^N(q^2) \nn
G_M^N(q^2) & = & F_1^N(q^2) + \kappa_N F_2^N(q^2) \quad.
\eqn
The direction of the (outgoing) photon momentum is chosen parallel
to the z-axis, and the initial nucleon is put in its rest frame.
Then, $G_E^N$ and $G_M^N$ can be calculated by
solving the set of equations for $j_0$ and $j_+ = \half (j_1 + i j_2)$.
Of course, the continuity equation has to be fulfilled,
which serves as a check of the calculation:
\bqn
\partial^\mu j_\mu = \partial^0 j_0 + \partial^3 j_3 = 0
\eqn
In our framework the continuity equation follows directly from
($\mu = 0, 3$)
\bqn
\bramket{P' s'}{j_\mu}{P s} & = & (-1)^{1-s-s'}
\bramket{P -s}{j_\mu}{P' -s'} \nn
& = & (-1)^{1-s-s'} \bramket{P s}{j_\mu}{P' s'} \quad,
\eqn
where the transformation properties of
Eqs.\ref{qcurrent2},\ref{dcurrent} under time reversal and parity
transformation (App.\ref{trafos}) have been used.
Indeed, we find that the current is conserved numerically for any momentum
transfers.

\section{Results and discussion}
\label{results}

The parameter values used in our model are given in Tab.\ref{param}.
The parameter influencing the shape of the form factors mostly is the
diquark parameter $\lambda$ (Eq.\ref{diquarkff}). After determining
the other parameters, $\lambda$ is chosen to give a best quantitative
description of the proton electric form factor. The obtained value is
in agreement with the commonly used diquark size in the range $0.1-0.3$ fm
\cite{anselmino,anselminokroll,fred}.
The quark and diquark
masses are free parameters: the constituent quark mass is chosen to be
in a range adopted by most quark models, see e.g. \cite{weise}.
The spectator diquark mass is a little less than the sum of
two quarks \cite{anselmino,vogl,ishii}. These mass values agree
with a binding energy of the nucleon proposed by the work of Ishii et
al. \cite{ishii} of 50 to 150 MeV. Thus, the nucleon appears as a
loosely bound state of quarks.
As outlined in Sect.\ref{solution},
the oscillator parameter is given by the minimum of the $M(\alpha)$
curve, and the coupling $g_{dqq}$ is adjusted by fitting the minimum
to the nucleon mass of 939 MeV.
Note that the so obtained oscillator parameter $\alpha = 1.15 $ fm
coincides roughly with
the assumed radius of the nucleon of about 1 fm \cite{weise}.
Our calculation contains 6 radial
basis functions, which is a sufficient approximation to the solution
of Eq.\ref{salpeter2}. The convergence of the solution is depicted in
Fig.\ref{alphaplot}.

\begin{figure}
\begin{center}
\leavevmode
\epsfxsize=0.65\textwidth
\epsffile{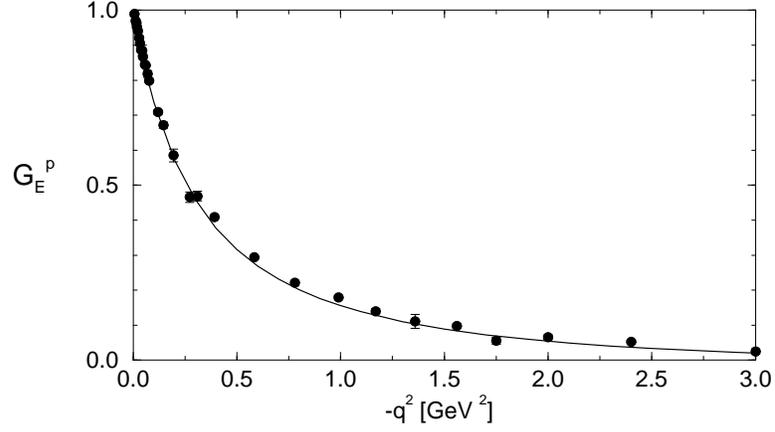}
\end{center}
\caption{The electric form factor of the proton $G_E^p(q^2)$;
  experimental data are taken from \protect\cite{hoehler}}
\label{geproton}
\end{figure}

\begin{figure}
\begin{center}
\leavevmode
\epsfxsize=0.65\textwidth
\epsffile{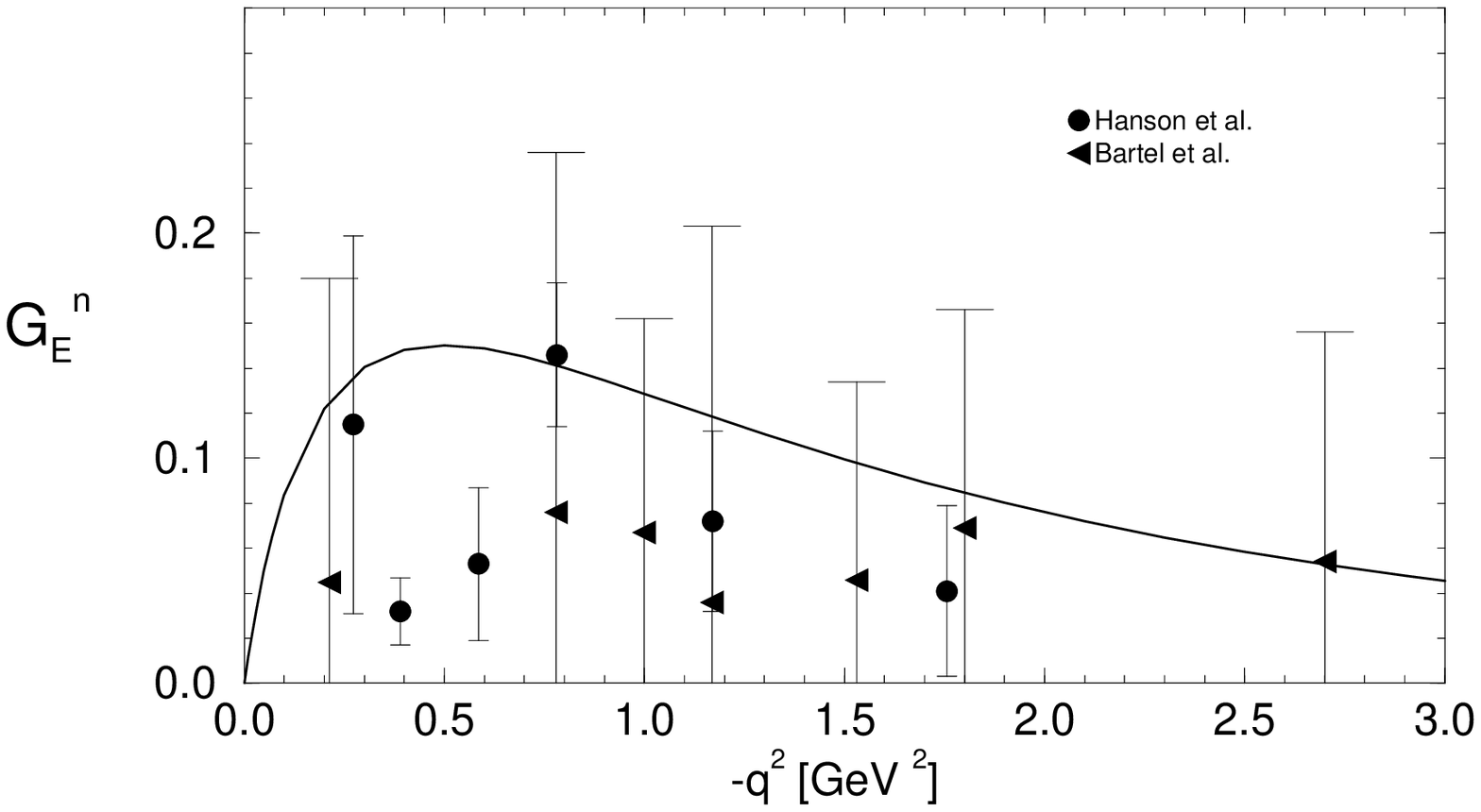}
\end{center}
\caption{The electric form factor of the neutron $G_E^n(q^2)$;
  experimental data are taken from \protect\cite{hanson,bartel}}
\label{geneutron}
\end{figure}

\begin{figure}
\begin{center}
\leavevmode
\epsfxsize=0.65\textwidth
\epsffile{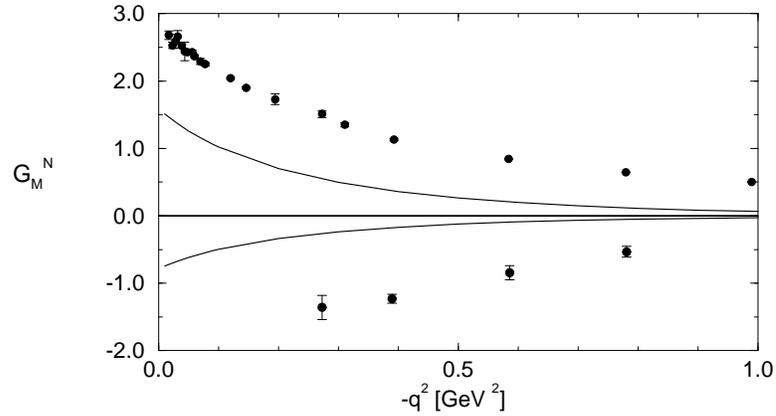}
\end{center}
\caption{The magnetic form factors $G_M^N(q^2)$ of the proton (upper
  curve) and of the neutron (lower curve);
  experimental data are taken from \protect\cite{hoehler,hanson}}
\label{gmnucleon}
\end{figure}

The calculated electric form factor of the proton $G_E^p$ as a
function of the negative photon momentum squared is shown in
Fig.\ref{geproton}. For $\lambda = 0.18$ fm  we find a very good
agreement with the
experimental data (taken from \cite{hoehler}). Even for $q^2$
as high as -3 GeV${}^2$ the calculation fits the data well.
{}From the slope of $G_E^p(q^2)$ at $q^2 = 0$ we obtain for the square
root of the mean square charge radius
\bqn
\sqrt{\langle r^2 \rangle_p} = \left(-6 \left.\frac{d G_E^p(q^2)}{d q^2}
                         \right|_{q^2 = 0} \right)^\half
 =  0.88 \; \mbox{fm} \quad,
\eqn
compared to the experimental value
$\sqrt{\langle r^2 \rangle_p} = (0.862 \pm 0.012) \; \mbox{fm}$
\cite{simon}.
The electric neutron form factor is compared to experiment
\cite{hanson,bartel} in Fig.\ref{geneutron}. We obtain a rather good
description of the
data, with a negative mean square charge radius
\bqn
\langle r^2 \rangle_n = - 0.28 \; \mbox{fm}^2 \quad,
\eqn
which is slightly larger than the experimental
$\langle r^2 \rangle_n = (- 0.117 \pm 0.002) \; \mbox{fm}^2$
\cite{dziembowski}.
The magnetic form factors of the proton and the neutron are shown in
Fig.\ref{gmnucleon}. Our results are a factor of about 2
too small. However, the calculation matches the experimental data
qualitatively. The shape of the curves is correct, and we obtain the
right signs. Of course, since $G_M$ is connected with the spin current,
effects of the vector component of a two-quark subsystem should be
important. Having neglected it, it is even more surprising to find
such a good correspondance.
Looking at $G_M(0)$, we obtain for the ratio
of the magnetic moments of neutron and proton
\bqn
\frac{\mu_n}{\mu_p} = - 0.5 \quad,
\eqn
compared to the experimental value $-0.685$. Of course, the above
ratio arises from the different charges of the single quark in the
nucleons; the scalar diquark does not contribute to the magnetic form
factor.
The prediction of the
right ratio $\frac{\mu_n}{\mu_p}$ is one of the successes of the SU(6)
quark model. Thus, it would be interesting to study the effects of an
additional axial-vector diquark component.
It should be noted that the neutron form
factors are related to the differences between the currents
\bqn
j_\mu^n \sim \frac{1}{3} \; j_\mu^{diquark}
       - \frac{1}{3} \; j_\mu^{quark} \quad.
\eqn
Thus, a description of the neutron form factors is most
sensible to the wave function.

\begin{figure}
\begin{center}
\leavevmode
\epsfxsize=0.65\textwidth
\epsffile{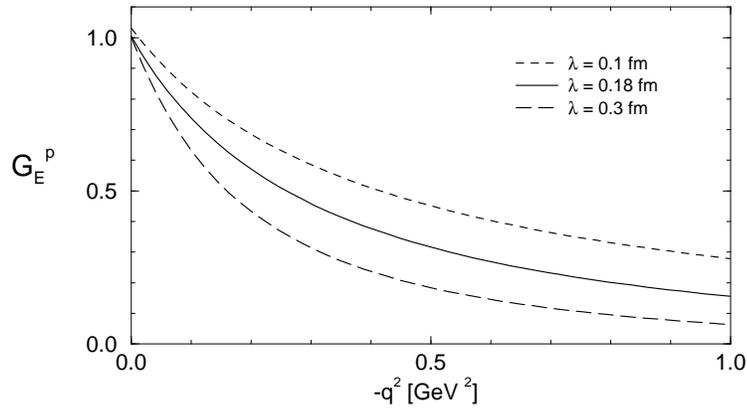}
\end{center}
\caption{The electric form factor of the proton $G_E^p(q^2)$
  for three different values of the diquark parameter $\lambda$}
\label{geplambda}
\end{figure}

\begin{figure}
\begin{center}
\leavevmode
\epsfxsize=0.65\textwidth
\epsffile{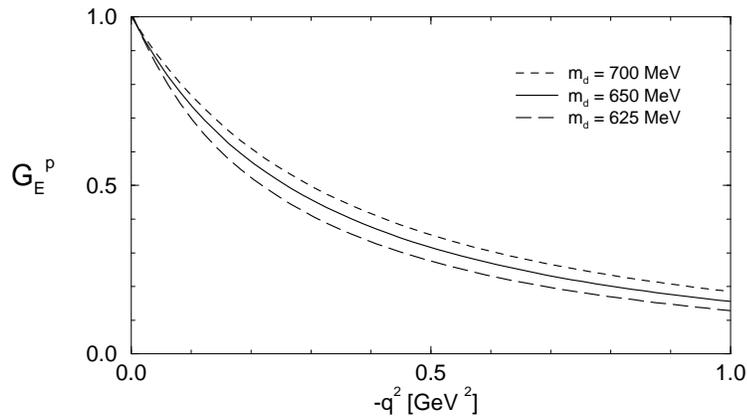}
\end{center}
\caption{The electric form factor of the proton $G_E^p(q^2)$
  for three different values of the diquark mass}
\label{gepdmass}
\end{figure}

In Fig.\ref{geplambda} we present the electric form factor of the
proton for three different diquark extensions $\lambda$.
The qualitative structure of the curve remains. Fig.\ref{gepdmass}
shows the variation of the electric form factor for three different
values of the diquark mass, with constant $\lambda = 0.18$
fm. Obviously, the variation is
small. However, we find that the diquark mass should not be smaller
than about 620 MeV: for $m_d = 600$ MeV the oscillator parameter is
$\alpha = 1.5$ fm, and rises sharply with decreasing $m_d$.

\section{Summary and outlook}
\label{summary}

We developed a simple, fully relativistic quark-diquark model of the
nucleon. In a first step, only scalar diquarks are taken into account.
Neglecting retardation effects of the interaction kernel, a Salpeter
equation for a bound state of a constituent quark and a scalar particle
is obtained. We assume a quark exchange as the only interaction.
The electric and magnetic form factors of the nucleon are then
calculated in a straightforward way using the Mandelstam formalism.
With four parameters (constituent quark and diquark masses,
diquark extension $\lambda$, diquark-quark coupling $g_{dqq}$,
see Tab.\ref{param}), we obtain
a very good description of the electric form factor up
to $q^2 = -3$ (GeV/c)$^2$. The neutron electric form factor is
reproduced almost quantitatively. The magnetic form factors are in
qualitative agreement with experiment. \\
Obviously, a relativistic treatment is extremely important to
describe observables of the nucleon in the medium energy range.
Furthermore, the assumption of a diquark structure of the nucleon
seems to be not out of place.

Of course, it would be interesting to study the effects of
eventual axial-vector diquarks. Especially for the magnetic form
factors they should be important. - The model will be extended to
baryons with strangeness. Then, various electromagnetic processes
may be studied. An example is the photoproduction of kaons. The
contributing amplitudes can be calculated in the framework of
the Mandelstam formalism, and a direct coupling of the photon
to the internal quark- and diquark lines is assumed (see \cite{vk}).

{\bf Acknowledgements:} I am grateful to H.R. Petry, B.C. Metsch and
especially to C.R. M\"unz for many helpful discussions. This work was
supported by the Deutsche Forschungsgemeinschaft.

\newpage

\begin{table}
\begin{tabular}{cccc}
$m_q$ & $m_d$ & $g_{dqq}$ & $\lambda$ \\
\hline
\quad 350 MeV/c$^2$ \quad &
\quad 650 MeV/c$^2$ \quad &
\quad 14.15 \quad &
\quad 1.8 fm \quad \\
\end{tabular}
\caption{The parameters of the model}
\label{param}
\end{table}
\newpage

\newpage

\begin{appendix}

\section{Transformation properties}
\label{trafos}

Under a Lorentz transformation $\Lambda$, the Bethe-Salpeter amplitude for a
bound state of a scalar particle and a fermion transforms like a
Dirac spinor
\bqn \label{lorentz}
\chi_P(p) = S_\Lambda^{-1} \chi_{\Lambda P}(\Lambda p) \quad.
\eqn
Using $S^F(\Lambda p) = S_\Lambda S^F(p) S_\Lambda^{-1}$ we obtain
\bqn
\Gamma_P(p) = S_\Lambda^{-1} \Gamma_{\Lambda P}(\Lambda p) \nn
\ov\Gamma_P(p) = \ov\Gamma_{\Lambda P}(\Lambda p) S_\Lambda
\eqn
Under time reversal (using the standard convention like e.g. in
\cite{itzykson}) we find
\bqn \label{time}
\chi_P^s(p) = (-1)^{S-s} \; (-\eps) \;
{}^t{\ov\chi_{\tilde P}}^{-s} (\tilde p)
\eqn
with $S=\half$ for the nucleon and
$\eps = i \gamma^1 \gamma^3\gamma^0$. \\
Under parity transformation we have
\bqn \label{parity}
\chi_P(p) = \pi_P \gamma^0 \chi_{\tilde P} (\tilde p) \quad,
\eqn
and under charge conjugation:
\bqn \label{charge}
\chi_P^{b_1 f_2}(p) = S_C \chi_P^{\ov b_1 \ov f_2}(-p)^*
\eqn
with $S_C = i \gamma^2$.
\end{appendix}

\end{document}